\documentclass[aps,prl,twocolumn,superscriptaddress,showpacs,
floatfix]{revtex4}
\usepackage{graphicx,epsfig}

\newcommand{\be}{\begin{equation}}
\newcommand{\ee}{\end{equation}}
\newcommand{\bary}{\begin{eqnarray}}
\newcommand{\eary}{\end{eqnarray}}
\newcommand{\ea}{E_{\alpha}}
\newcommand{\eb}{E_{\beta}}

\newcommand{\nn}{\nonumber\\}
\newcommand{\Del}{\Delta(0)}
\newcommand{\ua}{u_{\alpha}}
\newcommand{\va}{v_{\alpha}}
\newcommand{\ub}{u_\beta}
\newcommand{\vb}{v_\beta}

\newcommand{\mua}{\mu_\alpha}
\newcommand{\mub}{\mu_\beta}

\begin{document}

\title{Effective Vortex Mass from Microscopic Theory}
\author{Jung Hoon Han}
\affiliation{Department of Physics, Sung Kyun Kwan University, Suwon
440-746, Korea} \affiliation{CSCMR, Seoul National University, Seoul
151-747, Korea}
\author{June Seo Kim}
\affiliation{Department of Physics, Sung Kyun Kwan University, Suwon
440-746, Korea}
\author{Min Jae Kim}
\affiliation{Department of Physics, Sung Kyun Kwan University, Suwon
440-746, Korea}
\author{Ping Ao} \affiliation{Department of  Mechanical Engineering,
University of Washington, Seattle, WA 98195, USA}

\date{\today}

\begin{abstract}
We calculate the effective mass of a single quantized vortex in the
BCS superconductor at finite temperature. Based on effective action
approach, we arrive at the effective mass of a vortex as integral of
the spectral function $J(\omega)$ divided by $\omega^3$ over
frequency. The spectral function is given in terms of the
quantum-mechanical transition elements of the gradient of the
Hamiltonian between two Bogoliubov-deGennes (BdG) eigenstates. Based
on self-consistent numerical diagonalization of the BdG equation we
find that the effective mass per unit length of vortex at zero
temperature is of order $m (k_f \xi_0)^2$ ($k_f$=Fermi momentum,
$\xi_0$=coherence length), essentially equaling the electron mass
displaced within the coherence length from the vortex core.
Transitions between the core states are responsible for most of the
mass. The mass reaches a maximum value at $T\approx 0.5 T_c$ and
decreases continuously to zero at $T_c$.
\end{abstract}


\maketitle

Phase coherence is the defining characteristic of the superfluid
matter, and vortices - the quantized twist of the underlying phase
texture - are the unique elementary excitations of the condensate.
Their ubiquity as well as their role in transport and phase
transition makes the dynamics of an isolated vortex or their array an
integral part of our understanding of superfluidity.

The motion of a single vortex may be phrased as a Newtonian equation,
$ M_v   ( d^2 \mathbf{r}_v /dt^2 ) = \mathbf{F}_v$,  where $\mathbf{
F}_v$ is the force acting on the vortex, and $M_v$ is its effective
mass. The position of the vortex is represented by $\mathbf{r}_v$.
The mass of a vortex is \textit{entirely effective} in
nature\cite{wilczek}: it requires a force to drag a $h/2e$ flux
quantum across a type-II superconductor by an external magnet
(assuming zero mass for the magnet itself) because, in so moving, the
vortex interacts with the surrounding electrons. The virtual
transitions among the quasiparticle states caused by the vortex
motion lead to renormalization of mass, which in this case will
account for the entire vortex mass. Real transitions between the
states, on the other hand, lead to dissipation.

The Newtonian description of the vortex motion is a limiting case of
the more general quantum mechanical formulation. Quantum-mechanical
law of motion is most easily derived in the effective action
approach\cite{cl}. In essence, one divides the dynamics into that of
quasiparticles and the vortex, then by integrating out the
quasiparticle degrees of freedom, one is left with the effective
dynamics of the vortex alone. This way of studying vortex dynamics
was first conceived in the work of \u{S}im\'{a}nek\cite{simanek} and
later extensively developed by Ao and Zhu\cite{az1,az2}. In this
paper, we derive the mass formula from the effective action and
calculate it using BCS theory for temperatures $0 < T < T_c$.

The effective action $S_{eff}$ of a single vortex of a BCS
superconductor, centered at $\mathbf{r}_v (\tau)$ at imaginary time
$\tau$, is given by\cite{simanek,az1,az2}

\be {1\over 8}\int_0^\beta \! d\tau \int_{-\infty}^{\infty}\!
d\tau^\prime \int_0^\infty d\omega J(\omega) e^{-\omega
|\tau\!-\!\tau^\prime |}|\mathbf{r}_v (\tau)\!\!-\!\!\mathbf{r}_v
(\tau')|^2. \label{eq:eff_action}\ee  The spectral function is the
quantity \be J(\omega)\!=\!\sum_{ab}\delta
(\omega\!\!-\!\!|E_a\!\!-\!\!E_b |) |f(E_a )\!\!-\!\!f(E_b )|
\left|\langle a| {\partial H_0 \over
\partial \mathbf{r}_v} |b \rangle \right|^2.
\label{eq:spectralJ}\ee Fermi distribution function with energy $x$
is denoted $f(x)$. There is an additional action in $S_{eff}$
pertaining to the transverse motion of the vortex, which we do not
consider here\cite{az2}. The states $a,b$ are the eigenstates of the
Bogoliubov-deGennes (BdG) Hamiltonian $H_0$ in the presence of vortex
at $\mathbf{r}_v$, with energies $E_a $ and $E_b$ respectively. When
we assume that the dynamics is sufficiently \textit{ local in
time}\cite{comment}, we may approximate $\mathbf{r}_v
(\tau)\!\!-\!\!\mathbf{r}_v (\tau')\!\approx\!(\tau\!-\!\tau' )
\dot{\mathbf{r}}_v (\tau)$ and write

\be S_{eff} \approx {1\over2}\times\left(\int_0^\infty d\omega
{J(\omega) \over{\omega^3}}\right)\times \int_0^\beta d\tau
\left({{d\mathbf{r}_v}\over d\tau}\right)^2. \label{eq:s_eff} \ee The
effective vortex mass is the quantity in parenthesis, given by

\be M_v = \sum_{ab}\left| {{f(E_a )-f(E_b )} \over {(E_a \!-\!E_b
)^3}}\right| \left|\langle a| {\partial H_0 \over
\partial \mathbf{r}_v} |b \rangle \right|^2. \label{m-eff} \ee
This formula allows an explicit calculation of the vortex mass using
the BCS theory of superconductivity at arbitrary temperature.

Conclusions in the past vary regarding the vortex effective mass. One
group of theories predicts a ``small" mass of roughly one electron
mass per atomic length\cite{small-mass}, whereas another group
predicts a ``large" mass, of order $m (k_f \xi_0 )^2$ ($k_f$= Fermi
momentum, $\xi_0$ = coherence length at $T=0$)\cite{big-mass}. In
either case the theories are limited to zero temperature, and
finite-temperature generalization does not seem to be
straightforward.

The eigenstates of the BdG equation in the cylindrical coordinates is
written in the general form\cite{deGennes}

\be \psi_a (\mathbf{r} ) = {1\over\sqrt{2\pi L}}e^{ik_z z}e^{i\mu
\theta} \left(
\begin{array}{c}
  e^{-i{\theta\over 2}}u_\alpha (r) \\
  e^{+i{\theta\over 2}}v_\alpha (r)
\end{array} \right)\label{general-form}
\ee in the presence of a gap function $\Delta(\mathbf{r}) =
\Delta(r)e^{-i\theta}$ for a rectilinear vortex of length $L$
centered at $r=0$. The radial functions $(\ua (r) ,\va (r))$ are
obtained as eigenfunctions of the coupled differential equation ($r$
dependence in $u,v$ and $\Delta$ is implicit)

\bary   r^2 u^{\prime\prime}\!+\! ru^\prime \!+\![(k_r^2 \!+\!2m
E)r^2\!-\!\left(\mu\!-\!{1\over2}\right)^2]u
 &=& 2m r^2 \Delta v  \nn
r^2 v^{\prime\prime}\!+\!rv^\prime \!+\![(k_r^2\! -\!2m
E)r^2\!-\!\left(\mu\!+\!{1\over2}\right)^2]v
 &=& -2m r^2 \Delta u,
\label{eq:BdG_eq} \eary where $k_r^2=k_f^2-k_z^2$. Self-consistency
requires that the $r$-dependent gap function obey the relation
$\Delta(r)= V \sum_\alpha \ua (r) \va(r)[1-2f(\ea)]$ for some choice
of the pairing interaction strength $V$. The eigenstates are labeled
by $k_z$, $\mu$, and $\alpha$, which labels a set of energy levels
for a given choice of $(k_z, \mu)$. Uniqueness of the wavefunction
requires that $\mu$ be half-odd integers. Using these eigenstates,
the transition amplitude needed in evaluating Eq. (\ref{m-eff}) can
be written out in the form $-\langle \alpha|
\partial H_0 /\partial \mathbf{r}_v |\beta\rangle=
(\hat{x}\!\mp\! i\hat{y}/2)\times$

\be  \int_0^R \left\{ (\ua  \vb \!+\!\ub  \va  )r {{d\Delta  }\over
dr}\!\pm\! (\ua \vb  \!-\!\ub  \va  ) \Delta  \right\} dr .
\label{mat-ele1}\ee Upper and lower signs correspond to $\mu_\alpha
=\mu_\beta \pm 1$. The two eigenstates differ by one unit of angular
momentum due to the gradient operator which connects them. By
integrating by parts, Eq. (\ref{mat-ele1}) is equal to (assuming
$\mua =\mub +1 $)\begin{widetext} \be (E_\beta \!-\!E_\alpha
)\int_0^R \left\{ \left(\ua {{d\ub}\over dr}\!+\!\va{{d\vb}\over
dr}\right)r - \left((\mu_\beta\! -\!{1\over2})\ua \ub \!+\!
(\mu_\beta \!+\! {1\over2}) \va\vb\right)\right\} dr  + {R\over 2m}
\left({d \va \over dr}{d \vb \over dr} - {d \ua \over dr} {d \ub
\over dr}\right)_{r=R}.\label{mat-ele2}\ee
\end{widetext} We assume a hard-wall boundary condition at $r=R$.
One must be careful in treating the boundary term at $r=R$ on the
r.h.s. of the above equation, which is generally non-zero.

We follow earlier works\cite{numerics} and write the eigenfunctions
in the form \bary && \ua (r) = \sum_i c_{\alpha i} \phi_{ni} (r), ~~
\va (r) = \sum_i d_{\alpha i} \phi_{n+1,i} (r), \nn && ~~~~~\phi_{ni}
(r) = {\sqrt{2} \over R J_{n+1}(z_{ni})} J_n \left(z_{ni} {r\over
R}\right).\eary $z_{ni}$ is the $i$-th zero of the Bessel function
$J_n$. Integer value $n$ is related to the angular momentum by $\mua
= n+{1\over2}$. Coefficients $(c_{\alpha i}, d_{\alpha i})$ are
determined from matrix diagonalization. Negative-$\mu$ states need
not be considered separately, instead one can use a positive-$\mu$
eigenstate $(\ua,\va)$ of energy $\ea$ to construct a negative-$\mu$
eigenstate, given by $(\va,-\ua)$, of opposite energy $-\ea$.

With the eigenfunctions thus obtained, we evaluate the vortex mass
using Eq. (\ref{m-eff}), and the transition amplitude given by Eq.
(\ref{mat-ele2}). Furthermore, Eq. (\ref{mat-ele2}) can be
re-expressed using the coefficients $(c_{\alpha i}, d_{\alpha i})$
and making use of  Bessel identities. For $\mua=n+{1\over2}$,
$\mub=n-{1\over2}$, $n>0$, Eq. (\ref{mat-ele2}) becomes

\begin{widetext}
\be A_{n,\alpha\beta}={2\over R}\sum_{ij} c_{\alpha i}c_{\beta j}
z_{ni} z_{n-1,j} \left( {E_\alpha -E_\beta \over z_{ni}^2 -
z_{n-1,j}^2 }-{\xi_0 \over R^2} \right)+{2\over R}\sum_{ij} d_{\alpha
i}d_{\beta j} z_{n+1,i} z_{nj} \left( {E_\alpha -E_\beta \over
z_{n+1,i}^2 - z_{nj}^2 }+{\xi_0 \over R^2} \right),\label{term-A}\ee

and for $\mua=+{1\over2}$, $\mub=-{1\over2}$, it equals \be
B_{\alpha\beta}={2\over R}\sum_{ij} c_{\alpha i}d_{\beta j} z_{0i}
z_{1j} {J_0 (z_{1j}) \over J_2 (z_{1j}) }\left( {E_\alpha +E_\beta
\over z_{0i}^2 - z_{1j}^2 }-{\xi_0 \over  R^2} \right)+{2\over
R}\sum_{ij} d_{\alpha i}c_{\beta j} z_{1i} z_{0j} {J_0 (z_{1i}) \over
J_2 (z_{1i}) }\left( {E_\alpha +E_\beta \over z_{0j}^2 - z_{i1}^2
}-{\xi_0 \over R^2} \right).\label{term-B}\ee \end{widetext} In
deriving Eqs. (\ref{term-A})-(\ref{term-B}) we expressed energy in
units of the zero-temperature gap value at large distance from the
core, denoted $\Del$, and the length in units of $k_f^{-1}$. In Eq.
(\ref{term-A}), $\alpha$ and $\beta$ refer to eigenstates with a
given angular momenta $\mua$ and $\mub$, each. We made use of the
mapping between negative-$\mu$ and positive-$\mu$ eigenstates to
express the result in Eq. (\ref{term-B}) solely in terms of the
$\mu=+{1\over2}$ eigenstates. The vortex mass can be calculated using
$M_v =$ \bary  2m \!\cdot\! {\xi_0 \over 2}\!\cdot\! 2
\sum_{n>0,\alpha\beta}\left| {{f( \ea)-f( \eb)} \over
{(\ea\!-\!\eb)^3}}\right| A_{n,\alpha\beta}^2 && \nn \!\!+ 2m \cdot
{\xi_0 \over 2}\sum_{n=0,\alpha\beta}\left| {{f( \ea)-f(- \eb)} \over
{(\ea\!+\!\eb)^3}}\right| B_{\alpha\beta}^2.&&
\label{numerical-form}\eary Factor $2$ multiplying the first term
reflects the negative-$\mu$ contributions which equals those of the
positive-$\mu$ transitions.

Equations (\ref{term-A})-(\ref{numerical-form}) complete the formal
derivation of the vortex mass formula in the superconductor. Using
Eqs. (\ref{term-A})-(\ref{term-B}) gives dramatic improvement in both
the time and accuracy of the calculation of the transition amplitude
over the brute-force numerical integration of Eq. (\ref{mat-ele1}) or
(\ref{mat-ele2}).

In the following we discuss the vortex mass calculated using Eq.
(\ref{numerical-form}), with the eigenstates obtained from
self-consistent numerical diagonalization of Eq.
(\ref{eq:BdG_eq}). We choose the coherence length $E_f /\Del
\equiv \xi_0 = 20$ and the radius of the boundary $R=100$. For
each angular momentum $\mu$, eigenstates with energies within $\pm
10 \Del$ were retained. This left us with about 90 eigenstates for
$\mu=1/2$, and a decreasing number of states for larger $\mu$.
Calculations were restricted to $k_z =0$ only. Varying $k_z$ will
lead to different effective Fermi momentum $k_r = \sqrt{k_f^2
-k_z^2}$ in Eq. (\ref{eq:BdG_eq}) and will not change the
qualitative conclusion from the $k_z =0$ case. The energy gap
vanishes completely at $T \approx 0.571\Del$, which we take as the
transition temperature $T_c$. We work at several (reduced)
temperatures $t\equiv T/T_c$ in the range $0 < t < 1$.

\begin{figure}
\includegraphics[scale=0.8]{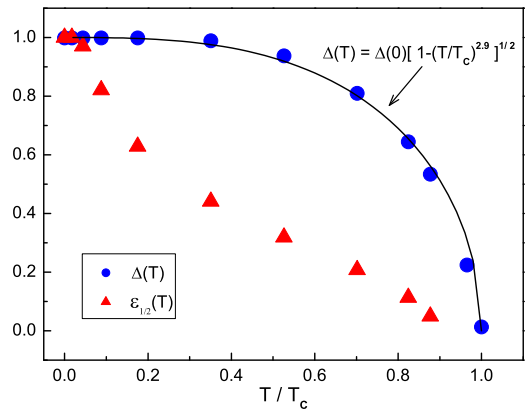}
\caption{Pair potential well away from the core, $\Delta(T)$ (blue),
and the core energy level for $\mu=+{1\over2}$, $\epsilon_{1/2}(T)$
(red), plotted vs. $T/T_c$. Both quantities are normalized by their
respective $T=0$ values, with $\epsilon_{1/2}(0)/\Del=0.088$ from
self-consistent calculation. $\Delta (T)$ shows excellent fit to
$\Del \sqrt{1-(T/T_c)^{2.9}}$ (black curve). } \label{fig1}
\end{figure}

Figure \ref{fig1} shows the calculated gap at large distance from
the core, $\Delta (T)$, and the $\mu=+{1\over2}$ core energy
level, $\epsilon_{1\over2}(T)$, for the temperatures we
considered. Both quantities decrease monotonically with $T$. For
$T$ sufficiently close to $T_c$ we can no longer resolve the core
levels as distinct from the continuum.

\begin{figure} \includegraphics[scale=1]{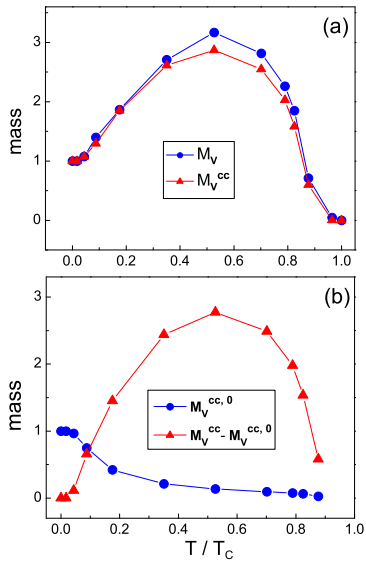}
\caption{(a) Effective mass $M_v (T)$ and the core-to-core
contribution to mass $M_v^{cc}(T)$ at temperature $T/T_c$,
normalized by $M_v (0)$. One finds $M_v (T) \approx M_v^{cc}(T)$
over most of $T$. (b) $ M_v^{cc}(T) = M_v^{cc,0}(T) +\sum_{n\neq
0} M_v^{cc,n}(T)$ (see text for definition). $M_v^{cc,0}(T)$ and
$M_v^{cc}(T)-M_v^{cc,0}(T)$ are plotted separately.
$M_v^{cc,0}(T)$ is monotonically decreasing while the higher-$n$
channels give a maximum mass at $T/T_c \approx 0.5$. The core
levels are not well resolved for $T$ too close to $T_c$, which
explains the absence of data points for $T\approx T_c$.}
\label{fig2}
\end{figure}

Referring to Eq. (\ref{numerical-form}) we may divide the mass as
arising from transitions between the core states ($M_v^{cc}$),
core-to-extended states ($M_v^{ce}$), and finally between extended
states ($M_v^{ee}$), as $M_v = M_v^{cc}+M_v^{ce}+M_v^{ee}$. As it
turns out the matrix element, Eq. (\ref{mat-ele1}), is vanishingly
small between a core and an extended state, and the mass is
effectively $M_v \approx M_v^{cc} + M_v^{ee}$. In Fig.
\ref{fig2}(a) we show the total mass $M_v (T)$ and the
core-to-core mass, $M_v^{cc}(T)$. As is evident from the figure,
$M_v^{cc}(T) \gg M_v^{ee}(T)$ for all temperatures except very
near $T_c$ where core levels are not resolved, but here the total
mass is vanishingly small anyway. Both the total mass $M_v$ and
the level-resolved masses $M_v^{cc}$ and $M_v^{ee}$ reach a
maximum value around $T=0.5T_c$.

The core-level contribution to mass can be further grouped
according to the angular momentum channels $(\mua
=n+{1\over2})\leftrightarrow (\mub=n-{1\over 2})$ over which the
transition takes place. We denote such angular-momentum-resolved,
core-to-core mass, $M_v^{cc,n}(T)$. Figure \ref{fig2}(b) shows
$M_v^{cc,0}(T)$ and $\sum_{n\neq 0}M_v^{cc,n}(T) =
M_v^{cc}(T)-M_v^{cc,0}(T)$. We find that $M_v^{cc,0}(T)$, between
$\mu =\pm {1\over2}$ core levels, survives at zero temperature and
monotonically decreases at higher $T$. The $n\neq 0$ channels give
zero mass for $T=0$, reaches a maximum at some intermediate $T$,
and decreases to zero at $T=T_c$. This behavior accounts for the
observed maximum in the mass $M_v (T)$.

The results shown in Fig. \ref{fig2}(b) can be nicely understood
thanks to the approximate identity that holds for the core
levels\cite{big-mass}, which states $|\langle \alpha |\nabla_v H_0
|\beta\rangle | \approx |\ea -\eb|$. We have verified that this
relation holds with excellent accuracy for all temperatures,
except when $T\approx T_c$. Then, we can approximate (see Eq.
(\ref{numerical-form})) \be M_v^{cc,n} (T) \approx 2m \xi_0 \left|
{{ f(\epsilon_{n-{1\over2}}(T) )-f(\epsilon_{n+{1\over2}}(T)) }
\over {\epsilon_{n-{1\over2}}(T)
\!-\!\epsilon_{n+{1\over2}}(T)}}\right|.
\label{approximate-core-mass}\ee In particular, for $n=0$ we have
the mass $M_v^{cc,0}(T)=2m \xi_0 \tanh (\beta \epsilon_{1\over2}
(T))/2\epsilon_{1\over2}(T)$. At zero temperature we get
$M_v^{cc,0}(0) = 2m\xi_0 /2\epsilon_{1\over2} (0)\sim m\xi_0^2$
since $\Del/\epsilon_{1\over2}(0)\sim \xi_0$. This is precisely
the mass of the electrons occupying the area $\sim \xi_0^2$. Using
the $\epsilon_{1\over2} (T)$ values shown in Fig. \ref{fig1}, we
find that the approximate formula in Eq.
(\ref{approximate-core-mass}) for $n=0$ reproduces $M_v^{cc,0}(T)$
obtained from the more general formula, Eq.
(\ref{numerical-form}), with very high accuracy.

On the other hand, $M_v^{cc,n\neq 0}(T)\approx\beta/ (4\cosh^2
(\beta \epsilon_n (T)/2))$, $\epsilon_n (T) \equiv
(\epsilon_{n+{1\over2}}(T)+\epsilon_{n-{1\over2}}(T))/2$, provided
the temperature is much larger than the typical core energy
spacing, $T \gg
|\epsilon_{n+{1\over2}}(T)\!-\!\epsilon_{n-{1\over2}}(T)|$. As the
typical core energy spacings are a few percent of the energy gap
$\Delta (T)$, this is not a very restrictive condition except near
$T=0$. Then one can easily check that this approximate form for
$M_v^{cc,n}(T)$ rises to a maximum value for $T\approx 0.69
\epsilon_n$ provided $\epsilon_n$ is treated as
temperature-independent. Therefore, the maximum in the mass shown
in Fig. \ref{fig2}(a) is due to the core-to-core transitions in
the higher angular momentum channels, $\sum_{n\neq
0}M_v^{cc,n}(T)$.

Summarizing our findings, (i) At zero temperature the effective
mass $M_v (T=0)$ is the mass of the electrons forming a cylinder
of radius $\xi_0$. It increases upon higher temperature, reaching
a maximum at $T\approx 0.5 T_c$, and vanishes at $T_c$. (ii) The
transition between the localized eigenstates forming the core
spectrum is mainly responsible for the effective mass. Hence, $M_v
(T) \approx M_v^{cc}(T)$. On the other hand, the dissipation
experienced by the moving vortex is due to the extended states as
it requires transition between states of the same energy (See
Refs. \cite{az1,az2} for a discussion of vortex friction).

Suppose now that the core levels got smeared due to the
impurities. If all higher angular momentum core levels except
$n=0$ disappeared, we would expect $M_v^{cc,n\neq 0}(T)$ is
effectively zero, and the mass behaves as $M_v (T) \approx
M_v^{cc,0} (T)$ showing the monotonic behavior of Fig.
\ref{fig2}(b). If all core levels disappeared, we will be left
with the extended states contributions which are shown to be quite
small. It is conceivable that in this case the mass becomes
``small", in agreement with predictions of Ref. \cite{small-mass}.

In closing we mention that studying the size ($R$) dependence may be
necessary to get a better estimate of the extended-states mass,
$M_v^{ee}(R)$. Since the core states are localized within a few
coherence lengths from the origin, our choice $R=10\xi_0$ is
presumably adequate to conclude $M_v^{cc}(R=100)\approx
M_v^{cc}(R=\infty)$. Extended states do not meet this requirement, at
least for ultra-clean systems with no impurity scattering to provide
the cutoff. The question of friction\cite{bardeen,az1,az2}, which
involves transitions between same-energy extended states, will also
require the study of $R$-dependence of the transition rates. We will
return to these issues in a future publication.


\begin{acknowledgments} HJH was supported by grant No.
R01-2002-000-00326-0 from the Basic Research Program of the Korea
Science \& Engineering Foundation. We thank helpful conversation with
X.-M. Zhu.
\end{acknowledgments}


\begin{thebibliography}{99}
\bibitem{comment} This will be the case when the vortex is moving
sufficiently slowly, or if it's moving along a straight line.
\end{thebibliography}
\end{document}